# Strain-induced stabilization of Al functionalization in graphene oxide nanosheet for enhanced $NH_3$ storage

Yunguo Li[1], Abir De Sarkar[1,2a], Biswarup Pathak[2] and Rajeev Ahuja[1,2]

[1]Applied Materials Physics, Department of Materials Science and Engineering,
Royal Institute of Technology (KTH),
S-100 44 Stockholm, Sweden

[2]Condensed Matter Theory Group
Department of Physics and Astronomy, Box 516, Uppsala University
S-75120 Uppsala, Sweden

## Abstract

Strain effects on the stabilization of Al dopant atom in graphene oxide (GO) nanosheet as well as its implications for $NH_3$ storage have been investigated using first-principles calculations. The high binding energy of Al ad-atom on GO is found to be a false indicator of its stability. The structural instability clearly contradicts the otherwise high stability indicated by the energetics of Al's binding to GO, stressing the need to assess the stability both from energetic and structural perspectives. Tensile strain effectively stabilizes the Al ad-atom on the GO nanosheet by strengthening the C-O bonds through an enhanced electronic charge transfer from C to O atoms. Interestingly, the strength of the C-O bonds is found to be the correct index for Al's stability. The trend for the variation of binding energy with strain is at odds with the systematic analysis of electronic structure based on bond lengths and atomic charges. At an optimal level of strain, the $NH_3$ storage capacity of Al-decorated GO is found to reach its maximum, where each Al ad-atom is able to bind up to 6 $NH_3$ molecules to itself with binding energies in the range of 0.2-1.1 eV. In contrast, the unstrained Al-doped GO is not found to bind even a single $NH_3$ molecule.

[a] Corresponding Author (Electronic mail): abir.desarkar@physics.uu.se

# 1 Introduction

Carbon nanostructures are very promising candidates for multifarious applications in nanoscience and nanotechnology, due to their light weight, low cost and high surface to volume ratio[1-8]. However, carbon nanostructures are chemically inert in their pristine form, as gas molecules bind very loosely to them. This precludes most of its practical applications. Decorating or functionalizing these nanostructures with metal atoms are common and practical avenues to enhance their gas adsorption capacities[5, 7, 9-11], which is brought about by the strong Lewis acid-base interaction between metal atoms and gas molecules[1, 4, 12-13]. Functionalization enables myriad potential applications of carbon nanostructures, including catalysis, molecular sensors, nanotechnology, biotechnology, energy storage and biomedicine[1, 14-16]. For a maximal utility of the functionalization, it needs to meet these two ends concurrently: (i) *stability*: the binding of the dopant to the nanomaterial needs to be stable & (ii) *efficiency*: the functionalization is intended to be as effective as possible[17]. Weak binding of metal atoms to the substrates are still frequently encountered[14-15, 18], which brings down the utility of the functionalization. Several pathways have been explored to circumvent this problem, among which the employment of mechanical strain is found to be most economic and practical[6, 19-20]. Strain occurs naturally at the interfaces of nanomaterials that differ in their lattice constants, for instance, in the growth of Ge thin films on Si substrate. Besides, strain is applied in nano-electromechanical systems (NEMS) and nano-optomechanical systems (NOMS) to tune the electronic and optical properties of nanomaterials. Up to 15% mechanical tensile strain has been experimentally realized in graphene[21-22]. 10% tensile strain has been found to significantly enhance the binding of metal ad-atoms to graphene and in turn, the adsorption strength of the gas molecules on the metal ad-atoms. Zhou *et al.*[6] have subjected metal-decorated graphene to mechanical strain, which not only stabilized the metal ad-atoms and counteracted their clustering propensity but also considerably enhanced their hydrogen storage capacity.

Shevlin and Guo[19] introduced defects in carbon substrates via creation of vacancy to anchor metal dopants, while Zhou et al.[23] have demonstrated an improvement in $H_2$ molecular adsorption on a polarizable substrate by the application of electric field. Kim et al.[20] and Liu et al[24] have resorted to subsitutional doping of boron to facilitate a stronger binding of metal ad-atoms on carbon substrates. A large number of studies on stabilizing the functionalization of metal atoms on carbon nanostructures have advanced its scientific understanding. Yet, it still continues to be an ongoing issue, underscoring the need for a much deeper understanding and insight.

Among the metal functionalized carbon nanostructures, metal-decorated graphene oxide (GO) materials have been increasingly gaining popularity for its superior electronic properties, high surface to volume ratio and low cost[7, 25-31]. Actually, GO was already synthesized using a simple procedure way back in 1958[32]. In this article, we have explored the potential of Al-decorated GO as a $NH_3$ storage material, as Al-doped carbon nanostructures have been found to bind gas molecules strongly[33]. Moreover, Al-functionalized GO has not yet been reported.

Usually, binding energy is considered as the stability index of an ad-atom in a system. However, in certain cases, the incorporation of an ad-atom into the system is accompanied by a large number of structural changes. In such cases, it may turn out to be a false indicator of the stability, necessitating a close look into the structure in order to ascertain the actual stability. Al-decorated GO is one such system where the binding energy of Al misrepresents its stability and the structural details warrant a closer inspection. It is the purpose of this article to probe into the microscopic mechanism behind the strain-induced stabilization of Al ad-atom in the GO using first principles density functional theory and its implications for a more efficient storage of $NH_3$ molecules. The paper is organized into a few sections.

## 2 Computational Details

All calculations have been performed using the *ab initio* density functional theory as implemented in the Vienna ab initio simulation package (VASP)[34]. The Perdew-Burke-Ernzerhof (PBE)[35] variant of the generalized gradient approximation (GGA-PBE) has been used for the exchange and correlation functional within the projector-augmented wave (PAW)[36] framework. The system is modeled as a $3 \times 4$ supercell of graphene (24 C atoms). A vacuum thickness of 16 Å in the direction normal to the surface was used to decouple the periodic images. A $4 \times 3 \times 1$ Monkhorst-Pack[37] k-mesh was used to sample the Brillouin zone. An energy cut off of 700 eV was used for the expansion of plane-wave basis. Atomic positions were optimized until the maximum force on each atom reached less than 0.001 a.u.

## 3 Results and Discussion

For a systematic analysis and discussion based on different perspectives, this section is organized into different sub-sections in order to reveal the underlying microscopic mechanism in stabilizing the functionalization of Al in GO.

### 3.1 The structure of Al-decorated GO nanosheet

First of all, we have ascertained the optimum position of the Al ad-atom on GO by geometry optimization calculations. Fig. 1(a) shows the supercell containing 24 carbon atoms and two epoxy oxygen atoms. According to the previous studies on GO nanosheet, the most stable positions for oxygen atoms in GO are the bridge sites of two adjacent carbon atoms. The most stable proposed structure of GO sheet with only epoxy functional units is that the epoxies are arranged in a line but at a distance of 3 C-C bond length on both sides of the graphene sheet[7, 26]. So, we optimized the structure of GO with two epoxies as is shown in Fig. 1(a, b). The C-O distance is found to be 1.47 Å, and the bonds between functionalized carbon atoms lengthened from 1.42 Å (the C-C bond length in graphene) to 1.48 Å, which corresponds well with the previous reports[7, 25]. The charge on the O atom bonded to C is found to be -0.82 e. O being more electronegative than C draws electronic charge towards itself. Then, we placed an Al atom at the center of the two epoxies at a distance of 2.00 Å from the epoxy O atoms. Fig.1(c, d) show the stable structure obtained from the optimization calculations, in which the oxygen atoms cleave their bonds with one of the carbon atoms and strike a strong bond with the decorated Al atom. In the optimized structure of Al-doped GO, the C-O bond length increases to 1.49 Å, while the Al-O bond length reaches 1.75 Å. Two valence electrons of the Al ad-atom are transferred to the two O atoms on its either side. As a result, the electronic charge on Al ad-atom and O atom reaches +2.00 e and -1.5e respectively. The O atoms gain the balance of -0.5e electronic charge from the C atoms of the underlying nanosheet.

### 3.2 The binding of Al ad-atom to GO nanosheet

The binding energy of an Al ad-atom on to the GO nanosheet is calculated as:

$$E_{BE} = (E_{GO} + E_{Al}) - E_{(GO:Al)} \qquad (1),$$

where $E_{GO}$ is the total energy of GO, $E_{Al}$ is the total energy of Al atom, and $E_{(GO:Al)}$ is the total energy of relaxed Al-decorated GO. The binding energy comes out to be 5.20 eV. It is much higher than that of Al$_2$ dimer in vacuum (1.62 eV) and a single Al ad-atom on graphene sheet (1.05 eV), ruling out the possibility for clustering of Al atoms.

But, counter-intuitively, the binding energy does not turn out to be the right index of its stability in GO. Moreover, the calculated binding energy of Al on GO doesn't correctly represent the binding of Al to GO.

In Al-decorated GO nanosheet, each Al ad-atom is linked to the underlying nanosheet via two O atoms. Prior to decoration, these two O atoms were located at their respective epoxy positions, bonded to two underlying C atoms of the nanosheet.

In the decorating process, the Al ad-atom bonds to the two epoxy O atoms on its either side. As a result, each of these two epoxy O atoms snaps one of their O-C bonds. The O-C bonds that survived the process of Al-doping were slightly weakened as indicated by the increase in O-C bond length. If we ignore the weakening of these O-C bonds in Al-decorated GO, the bond strength of O-Al can be expressed as:

$$E_{BE(O-Al)} = ((E_{GO} + E_{Al}) - E_{(GO:Al)} + E_{BE(O-C)})/2 \qquad (2),$$

where $E_{GO}$, $E_{Al}$ and $E_{(GO:Al)}$ refer to the same energies defined in eqn. (1), and $E_{BE(O-C)}$ is the bond energy of the O-C bond in GO. Since $E_{BE(O-C)}$ is found to be 1.26 eV in our calculations, the bond energy of the O-Al bond, $E_{BE(O-Al)}$ is then 3.86 eV. The factor of ½ in eqn. (2) accounts for the fact that the Al ad-atom bonds with two O atoms on its either side.

The Al ad-atom binds strongly to the O atoms of GO, which shows up in its high binding energy. The strong binding of the Al atom to the O atoms adversely affects the strength of the bonds of the O atoms with the underlying C atoms (designated as C1) of the substrate. From the perspective of bond order conservation, the strong O-Al bonds significantly weaken the bonds of the O atoms with the C atoms of substrate. The O-C bonds are weakened to such an extent that the functional unit constituted by the Al atom and the two epoxy oxygen atoms is found to break its bond with the substrate sheet upon adsorption of a single NH$_3$ molecule. Here the situation is different from that of Al-decorated graphene sheet in the sense that the Al ad-atom is indirectly linked to the substrate via the O atoms of the oxide.

### 3.3 Al-decorated GO upon adsorption of a single NH$_3$ molecule

The Al-decorated GO has been investigated for the purpose of $NH_3$ storage. First, we have used Al-decorated GO for the adsorption of one $NH_3$ molecule, using different distances between the Al ad-atom and the $NH_3$ molecules as well as different orientations of the molecules as starting points for geometry optimizations. Unfortunately, when a single $NH_3$ molecule interacts with the Al-decorated GO, the $NH_3$ molecule is pulled to the Al ad-atom and the aforementioned O-Al-O functional unit rips off the substrate sheet, as shown in Fig.1 (e, f). It indicates that the adsorption of $NH_3$ further weakens the already weak O-C bonds, due to some bond order conservation. The O-C bonds are weakened to such an extent upon adsorption of a $NH_3$ molecule that the O-Al-O functional unit shown in Fig.1(c, d) dissociates itself from the underlying substrate sheet.

The striking features in Al-decorated GO are the weak O-C bonds. Actually, the Al-decorated GO may be looked upon as a graphene sheet decorated by the O-Al-O functional unit.

**3.4 Al-decorated GO under strain**

The tensile strain, varying from 0% to 15%, was applied along both arm-chair (x-axis) and zig-zag (y-axis) directions. The atomic positions were then relaxed in the strained structures. The Al-decorated GO was found to be stable up to 10% tensile strain along the arm-chair direction and simultaneously up to 15% along the zig-zag direction. The structure of the Al-decorated GO collapses when the applied strain exceeds this range. So we find the tolerance limit of Al-doped GO to tensile strain to be 10% along the arm-chair direction and 15% along the zig-zag direction.

An energetic analysis of the strain effects is not sufficient to reveal the underlying mechanism, as it involves a large number of changes in bond lengths in the system when it is subject to strain. In addition, we need to examine the strain-induced changes in the electronic structure and geometry of the O-Al-O functional unit. It is instructive to track the evolution in the binding of the O-Al-O functional unit to the underlying substrate under the application of mechanical strain.

Tensile strain induced effects show up more prominently through charge redistribution among the Al, O and C atoms than through changes in the bond lengths. The maximum changes in O-C and O-Al bond lengths are found to be 0.077 Å and 0.014 Å respectively for the range of strain investigated in our work. Tensile strain on

Al-decorated GO is found to shorten the bond length of O-C, while it lengthens the O-Al bond. Furthermore, strain along the armchair direction is found to be more effectual than along the zig-zag direction in strengthening the C-O bonds. So, the applied tensile strain is found to be very effective in stabilizing the Al ad-atom on GO by strengthening the O-C bonds, while simultaneously reducing the O-Al bonds by a relatively much smaller magnitude.

To further elucidate the electronic mechanism behind strain induced stabilization of Al ad-atom, a Bader analysis[38] of atomic charges was also performed. The charges on Al, O and the C (C1 carbon atom) bonded to O are shown in Tables 1, 2 and 3, respectively. Under the application of tensile strain, O gained charge mainly from C and marginally from Al, which is consistent with the changes in bond length. The applied tensile strain is found to redistribute electronic charge among the Al, O and the C1 atoms and thereby stabilize the functionalization of GO by Al. The O-C bond is strengthened by a significant charge transfer from C to O. In Table 4, the binding energy of Al ad-atom on GO (BE) shows an overall decreasing trend with the application of mechanical strain, implying a strain induced destabilization of Al functionalization in GO. This variation in the BE with strain is clearly found to be at odds with the systematic analysis of electronic structure based on bond lengths and atomic charge redistribution. It confirms that the BE is a false indicator of the stability of Al dopant atom in GO. The optimal strain is found to be 5% along the arm-chair direction and 15% along the zig-zag direction, where the O-C bond strength reached its peak. Fig.2 shows the iso-surfaces for the differential electronic charge density of both unstrained and optimally strained Al-decorated GO nanosheet. The differential charge density is obtained from:

$$\Delta\rho = \Delta\rho_{(GO:Al)} - (\Delta\rho_{Al} + \Delta\rho_{GO}) \quad (3),$$

where $\Delta\rho_{(GO:Al)}$, $\Delta\rho_{Al}$, and $\Delta\rho_{GO}$ denote the charge densities of the relaxed Al-decorated GO, Al atom and GO respectively. Accumulation and depletion of charge around O and C respectively is found to occur in significant proportions under the application of strain, while charge around Al depletes marginally under strain.

### 3.5 $NH_3$ adsorption on Al-decorated GO under strain

Thereafter, we have examined the adsorption of $NH_3$ molecules on the optimally strained Al-decorated GO nanosheet. Different distances between Al dopant and $NH_3$ molecules as well as different orientations of the molecules were used as starting points for geometry optimizations. The adsorption energy of NH$_3$ molecules and corresponding adsorbed structures are shown in Fig. 3.

To further investigate the energetics of NH$_3$ adsorption, the adsorption energy of NH$_3$ was calculated as

$$E_{ads,NH_3} = \frac{E_{GO:Al} + nE_{NH_3} - E_{GO:Al-NH_3}}{n} \qquad (4),$$

where n refers to the number of adsorbed NH$_3$ molecules, $E_{GO:Al}$ to the total energy of relaxed Al-decorated GO nanosheet. $E_{NH_3}$ is the total energy of a relaxed free NH$_3$ molecule, while $E_{GO:Al-NH_3}$ is the total energy of Al-decorated GO with adsorbed NH$_3$ molecule(s).

Al ad-atom binds the first NH$_3$ molecule with a significant decrease in the total energy of the system (1.25 eV), with the NH$_3$ molecule lying on one side of the plane of the O-Al-O functional unit. It is justified since Al loses considerable charge along the direction normal to the plane of the O-Al-O functional unit. The Al-N distance is 2.04 Å, while the O-Al and O-C bond lengths are 1.77 Å and 1.40 Å, respectively. The O-Al and O-C bond lengths remain almost unchanged upon adsorption, suggesting the stability of the functional unit against gas adsorption. The adsorption energy of the second NH$_3$ molecule is 1.53 eV, suggesting a strong adsorption for the second molecule as well. The two NH$_3$ molecules distributed themselves symmetrically on the two sides of the O-Al-O functional unit. The Li-N distances are 2.00 Å. The O-C distance slightly increased from to 1.43 Å, but the O-Al distance decreased to 1.72 Å upon adsorption of the second NH$_3$ molecule. It indicates that adsorption of the second molecule weakens the O-C bonds slightly and at the same time, it strengthens the O-Al bonds a little. For the third and fourth, the adsorption energies are found to be 0.44 eV and 0.37 eV and the Al-N distances are 3.49 Å and 3.56 Å, and the adsorption energy of the fifth and sixth molecules is 0.22 eV and 0.28 eV respectively. The O-C and O-Al bond lengths remain unchanged upon the adsorption of 3 to 6 NH$_3$ molecules. This shows that the adsorption of the molecules in succession hardly affect these two bonds. So, the adsorption of six NH$_3$ molecules is found to be attainable, and the adsorption energies are found to lie within a reasonable range. More importantly, the adsorption of NH$_3$ molecules has not been found to produce any discernible effect on the structure of the O-Al-O functional unit. Therefore, the applied strain is found to be very efficacious in stabilizing the structure, suiting it for $NH_3$ adsorption and storage.

## 4 Conclusions

In summary, using first-principle calculations, we have studied the strain effects on the stabilization of the O-Al-O functional unit on Al-decorated GO for $NH_3$ storage. The high binding energy of Al on the GO is found to be a false indicator of its stability. The structural instability of Al on GO is clearly at variance with the high binding energy of Al on GO. Therefore, it was found necessary to assess the stability both from energetic and structural perspectives. The applied mechanical strain is found to effectively stabilize the Al functionalization on GO, by tuning the charge redistribution among Al atom, O atom and the underlying graphene sheet. Tensile strain strengthens the C-O bonds; while it simultaneously weakens the O-Al bonds marginally due to bond order conservation. The strength of the C-O bonds is found to be the correct index for Al's stability in GO. While the unstrained Al-decorated GO nanosheet is unable to bind even a single $NH_3$ molecule, it is found to bind up to six $NH_3$ molecules with considerably good binding energies when it is strained optimally.

## Acknowledgments

Y.G. Li thanks the Chinese Scholarship Council (CSC), while A. De Sarkar, B. Pathak and R. Ahuja would like to gratefully acknowledge the funding support from the Wenner-Gren Stiftelserna/Foundation. R.A. also thanks FORMAS, STEM and SWECO, Sweden. SNIC and UPPMAX are acknowledged for computing time.

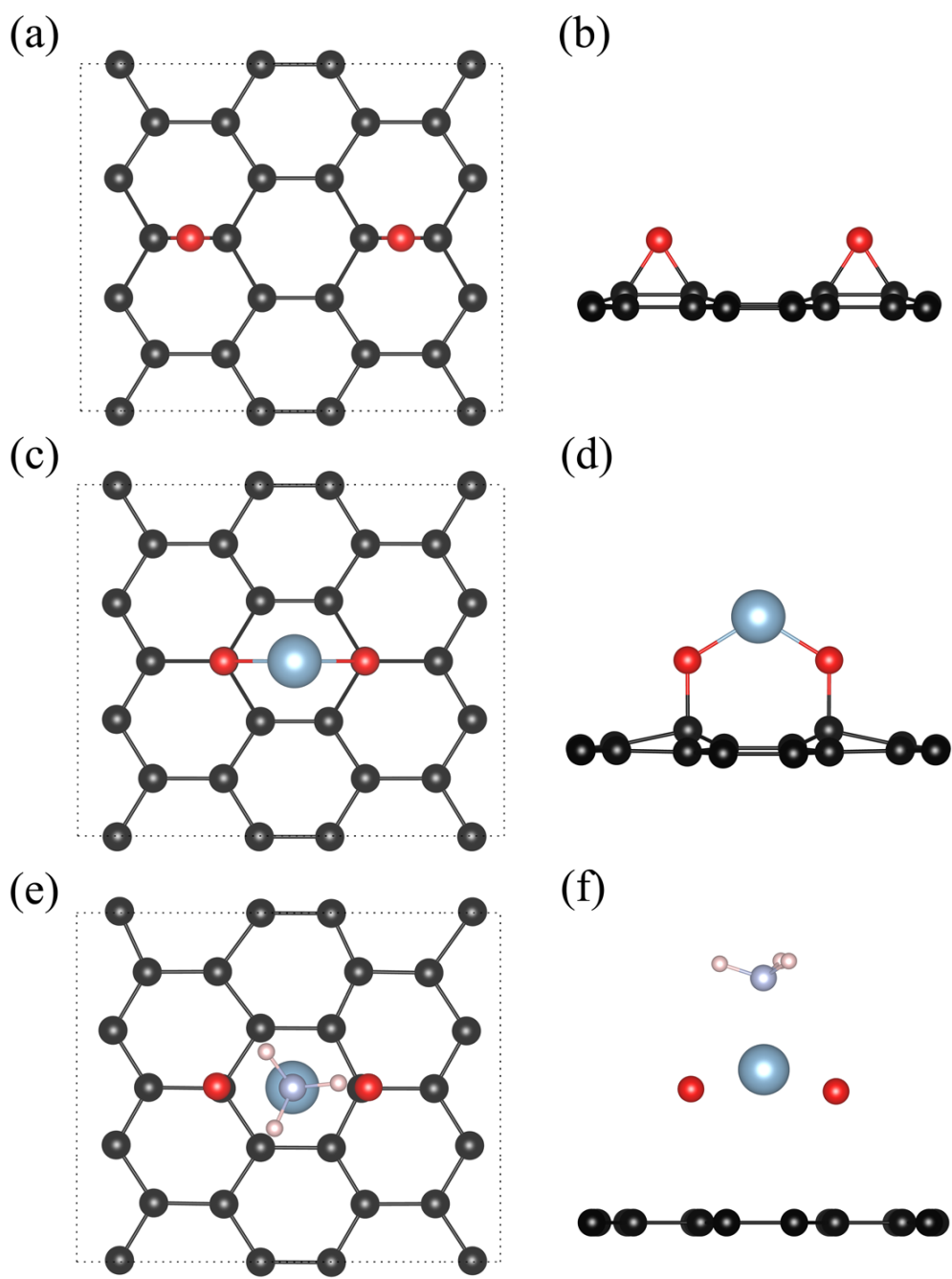

Fig. 1 (a) top view of graphene oxide (GO); (b) side view of GO; (c) top view of Al-decorated GO; (d) side view of Al-decorated GO; (e) top view of one $NH_3$ molecule on Al-decorated GO; (f) side view of one $NH_3$ molecule on Al-decorated GO

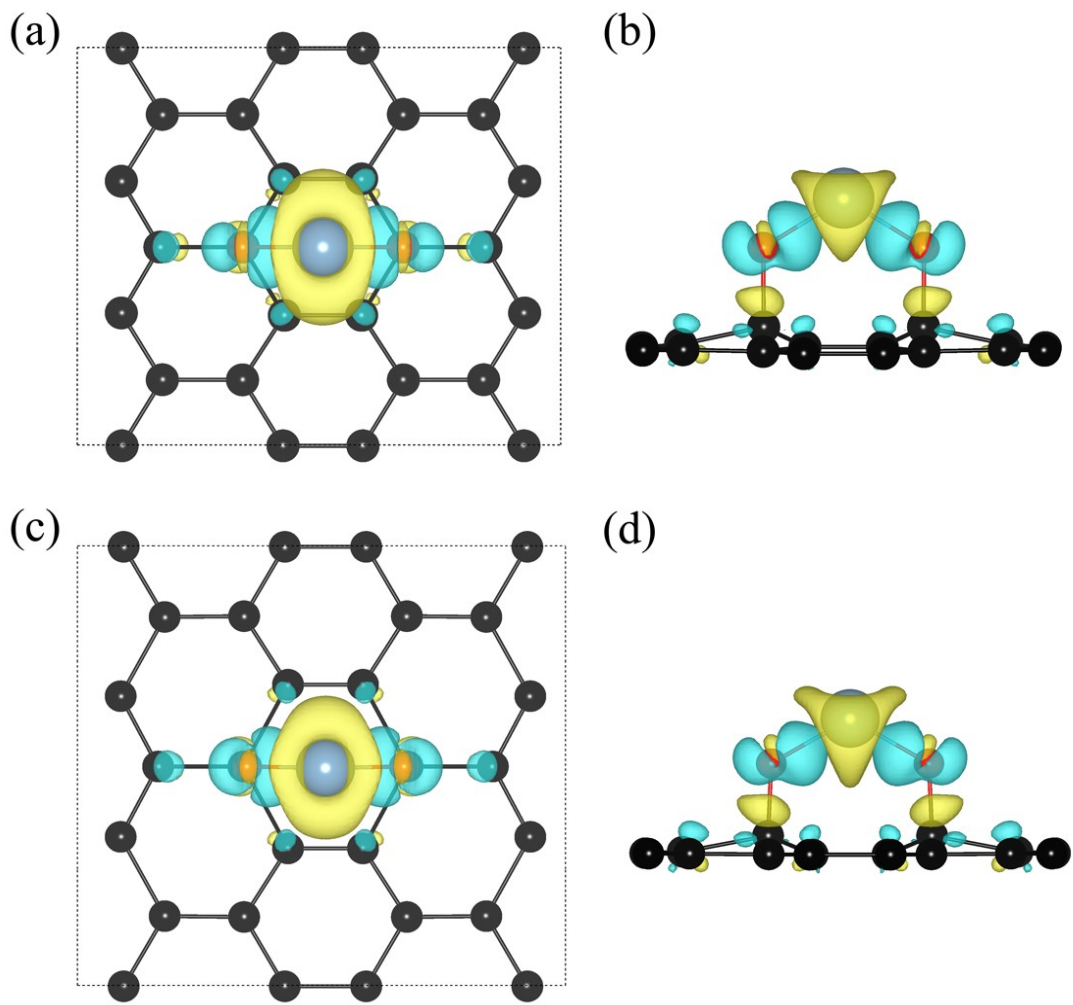

Fig. 2 Differential charge density plots (yellow: charge depletion, cyan: charge accumulation) (a) top view of unstrained Al-decorated GO; (b) side view of unstrained Al-decorated GO; (c) top view of optimally strained Al-decorated GO; (d) side view of optimally strained Al-decorated GO.

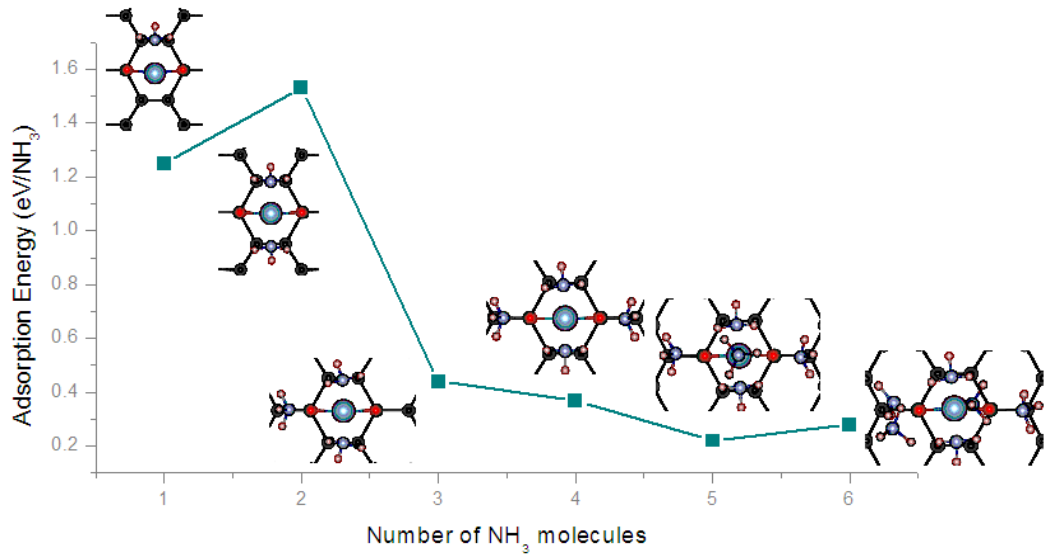

Fig.3 Binding/adsorption energy of $NH_3$ molecule(s) in eV on optimally strained Al-decorated GO.

Table 1. Electronic charge on Al ad-atom as a function of tensile strain applied along the two directions (X-axis: arm-chair, Y-axis: zig-zag)

|  | Strain along X-axis (%) | | | |
| --- | --- | --- | --- | --- |
| Strain along Y-axis (%) | 0 | 2.5 | 5.0 | 10 |
| 0 | +2.00 | +1.99 | +2.00 | +2.06 |
| 2.5 | +2.01 | +2.07 | +2.08 | +2.07 |
| 5.0 | +2.03 | +2.08 | +2.08 | +2.07 |
| 10 | +2.08 | +2.09 | +2.09 | +2.08 |
| 15 | +2.09 | +2.09 | +2.10 | —* |

* The structure has collapsed at this pair of strain values. The same applies to the tables below.

Table 2. Electronic charge on O atom as a function of tensile strain applied along the two directions (X-axis: arm-chair, Y-axis: zig-zag)

|  | Strain along X-axis (%) | | | |
| --- | --- | --- | --- | --- |
| Strain along Y-axis (%) | 0.0 | 2.5 | 5.0 | 10 |
| 0 | -1.52 | -1.69 | -1.72 | -1.77 |
| 2.5 | -1.74 | -1.77 | -1.8 | -1.89 |
| 5.0 | -1.76 | -1.79 | -1.84 | -1.89 |
| 10 | -1.83 | -1.88 | -1.93 | -2.03 |
| 15 | -1.94 | -1.97 | -2.04 | — |

Table 3. Electronic charge on one of the carbon atoms in the graphene sheet bonded to O under tensile strain applied along the two directions (X-axis: arm-chair, Y-axis: zig-zag). This carbon atom is designated as C1 atom in the text of the article.

|  | Strain along X-axis (%) | | | |
| --- | --- | --- | --- | --- |
| Strain along Y-axis (%) | 0.0 | 2.5 | 5.0 | 10 |
| 0.0 | +0.49 | +0.68 | +0.70 | +0.75 |
| 2.5 | +0.73 | +0.42 | +0.78 | +0.92 |
| 5.0 | +0.76 | +0.43 | +0.84 | +0.92 |
| 10 | +0.82 | +0.87 | +0.93 | +1.05 |
| 15 | +0.95 | +1.00 | +1.08 | — |

Table 4. Binding energy (BE) of the Al ad-atom to the GO nanosheet (in eV) under tensile strain applied along the two directions (X-axis: arm-chair, Y-axis: zig-zag)

| Strain along Y-axis (%) | Strain along X-axis (%) | | | |
|---|---|---|---|---|
| | 0 | 2.5 | 5.0 | 10 |
| 0 | 5.20 | 5.06 | 4.91 | 4.45 |
| 2.5 | 3.20 | 4.42 | 4.22 | 1.46 |
| 5.0 | 2.93 | 4.3 | 4.11 | 1.54 |
| 10 | 2.67 | 4.07 | 3.93 | 1.85 |
| 15 | 0.63 | 3.86 | 3.78 | — |